\documentclass[10pt,twocolumn]{article}

\usepackage[a4paper,top=1.8cm,bottom=2.0cm,left=1.6cm,right=1.6cm,columnsep=0.65cm]{geometry}
\usepackage[T1]{fontenc}
\usepackage[utf8]{inputenc}
\usepackage{times}
\usepackage{graphicx}
\usepackage{amsmath}
\usepackage{amssymb}
\usepackage{microtype}
\usepackage{caption}
\usepackage{cite}
\usepackage[hidelinks]{hyperref}
\usepackage{url}

\newcommand{\pT}{\ensuremath{p_{\mathrm{T}}}}
\newcommand{\snn}{\ensuremath{\sqrt{s_{\mathrm{NN}}}}}
\newcommand{\sqrts}{\ensuremath{\sqrt{s}}}
\newcommand{\HeThree}{\ensuremath{^{3}\mathrm{He}}}
\newcommand{\Lambdastar}{\ensuremath{\Lambda(1520)}}

\begin{document}

\twocolumn[
\begin{center}
{GENERAL ARTICLES}\\\vspace{1.2em}
{\LARGE\bfseries Why Do Light Nuclei Survive at the Large Hadron Collider?}\\\vspace{0.8em}
{\large Sushanta Tripathy and Raghunath Sahoo}\\\vspace{1em}
\end{center}

\noindent\textbf{Keywords:} Heavy-ion collisions, quark-gluon plasma, light nuclei, dark matter, Large Hadron Collider

\vspace{0.7em}
\noindent\textbf{Abstract:} Light nuclei and antinuclei, such as deuterons, are produced abundantly at the Large Hadron Collider (LHC) in hadronic and nuclear collisions. Even though their binding energies are only a few MeV, they survive in the extremely high temperatures of the order of a few hundred MeV. This contradiction, often referred to as ``Snowballs in Hell'', has become a sharp test of how quantum chromodynamics (QCD) turns quarks and gluons into composite matter. Strikingly, two very different frameworks can reproduce the same inclusive yields, i.e., late-stage nucleon coalescence, where nuclei form from nearby nucleons as the system dilutes, and statistical thermal models, where nuclei emerge as part of an equilibrated hadronization chemistry at a temperature close to 155 MeV. Here, we review how recent LHC measurements and model developments are shifting the question--from whether light nuclei are produced, to when and how they form, with broader implications for QCD matter and cosmic-ray antinuclei searches.

\vspace{1.5em}
]

\section*{Introduction}

The Large Hadron Collider (LHC) at CERN, Geneva, is the world's largest accelerator operating at the energy and luminosity frontiers. A major part of the LHC physics programme is devoted to our understanding of quantum chromodynamics (QCD), the theory that binds the fundamental building blocks of matter --quarks and gluons-- into hadrons \cite{star2010,alice2016a,sun2019}. Hadronization, the process of formation of hadrons from quarks and gluons, is an early-stage phenomenon in the space-time evolution of the produced fireballs in proton and heavy-ion collisions at GeV/TeV energies. During the hadronization process, coloured quarks and gluons become colour-neutral hadrons, which lie in the strong-coupling regime and hence are difficult to access directly in experiments. Here, controlled theoretical calculations are limited, and precision experimental data often dictates which physical pictures remain feasible.

Light nuclei have become an unusually intriguing test case. These are not ``exotic'' in a particle-physics sense; they are nuclear bound states, but produced in an extreme, rapidly expanding environment. The central question is simple to state but hard to answer: how do weakly bound composite objects emerge from a system whose characteristic energy scales are orders of magnitude larger than nuclear binding energies? For example, the deuteron binding energy is only about 2.2 MeV, yet such states appear in both Pb--Pb and even proton-proton (pp) collisions, where the system size and lifetime are significantly different, with system temperatures ranging from around 350 MeV at the early stage of the created fireball to around 90 MeV at kinetic freeze-out, when the produced secondary particles free-stream to reach the detectors \cite{alice2016b,alice2013}.

Understanding how light nuclei form in high-energy collisions is not only a long-standing puzzle in collider physics, but also crucial for astrophysics. Space-borne experiments such as AMS-02 on the International Space Station are searching for rare signals of antideuterons and antihelium that could be linked to indirect dark-matter searches \cite{korsmeier2018,blum2017}. At the same time, the same antinuclei can be produced as secondaries in ordinary high-energy interactions of cosmic rays with the interstellar medium, creating an unavoidable background. LHC measurements help by pinning down the production mechanisms and reducing uncertainties in the background estimates, thereby strengthening the interpretation of any future antinuclei candidates.

\section*{Nuclei puzzle: coalescence vs thermal production}

\begin{figure}[t]
  \centering
  \includegraphics[width=0.88\columnwidth]{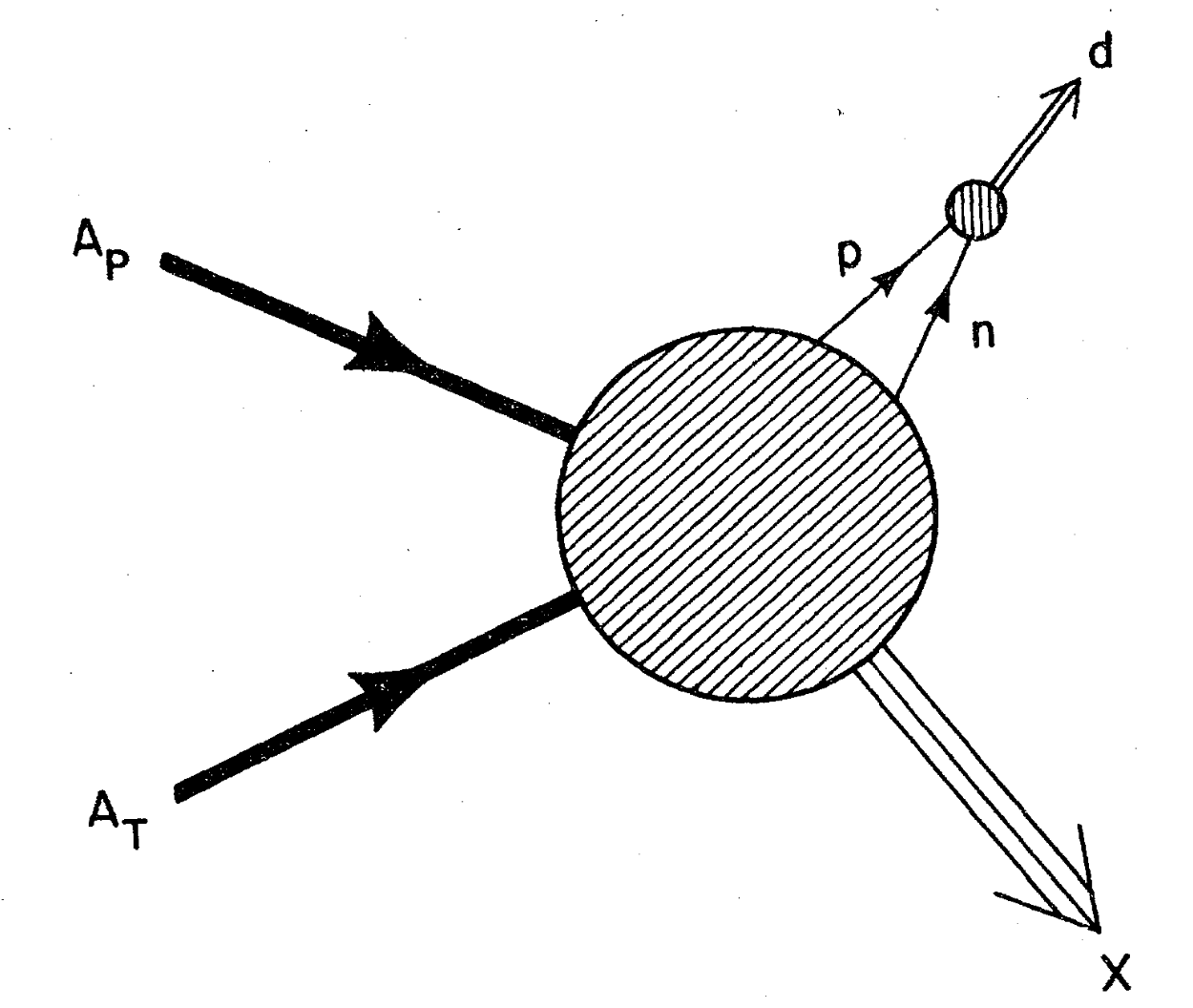}
  \caption{Depiction of deuteron production via coalescence \cite{kapusta1980}.}
  \label{fig:coalescence}
\end{figure}

\begin{figure}[t]
  \centering
  \includegraphics[width=0.98\columnwidth]{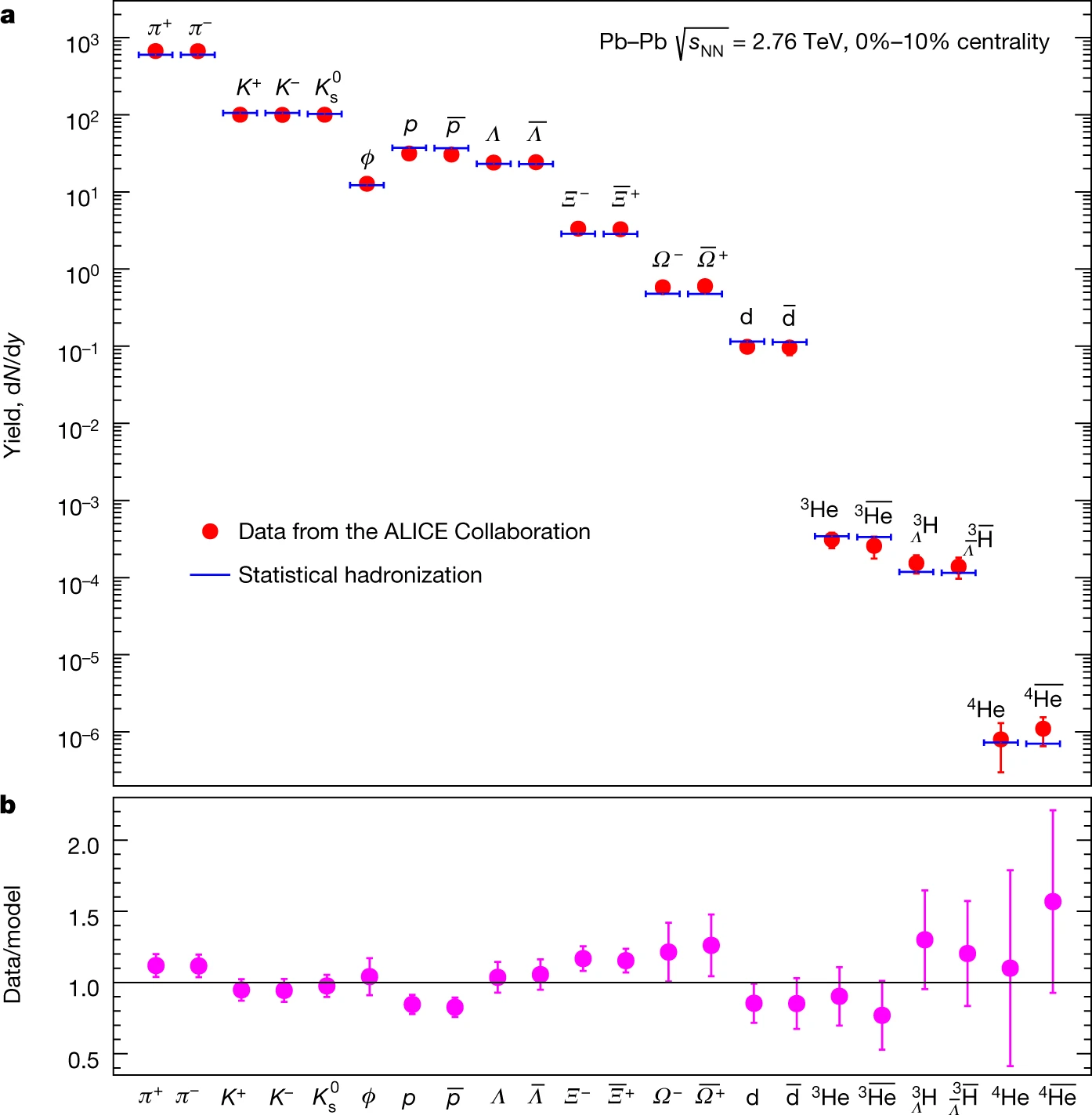}
  \caption{Hadron abundances in Pb--Pb collisions at \snn{} = 2.76 TeV compared with statistical hadronization model predictions \cite{andronic2018}.}
  \label{fig:shm}
\end{figure}

\begin{figure}[t]
  \centering
  \includegraphics[width=0.98\columnwidth]{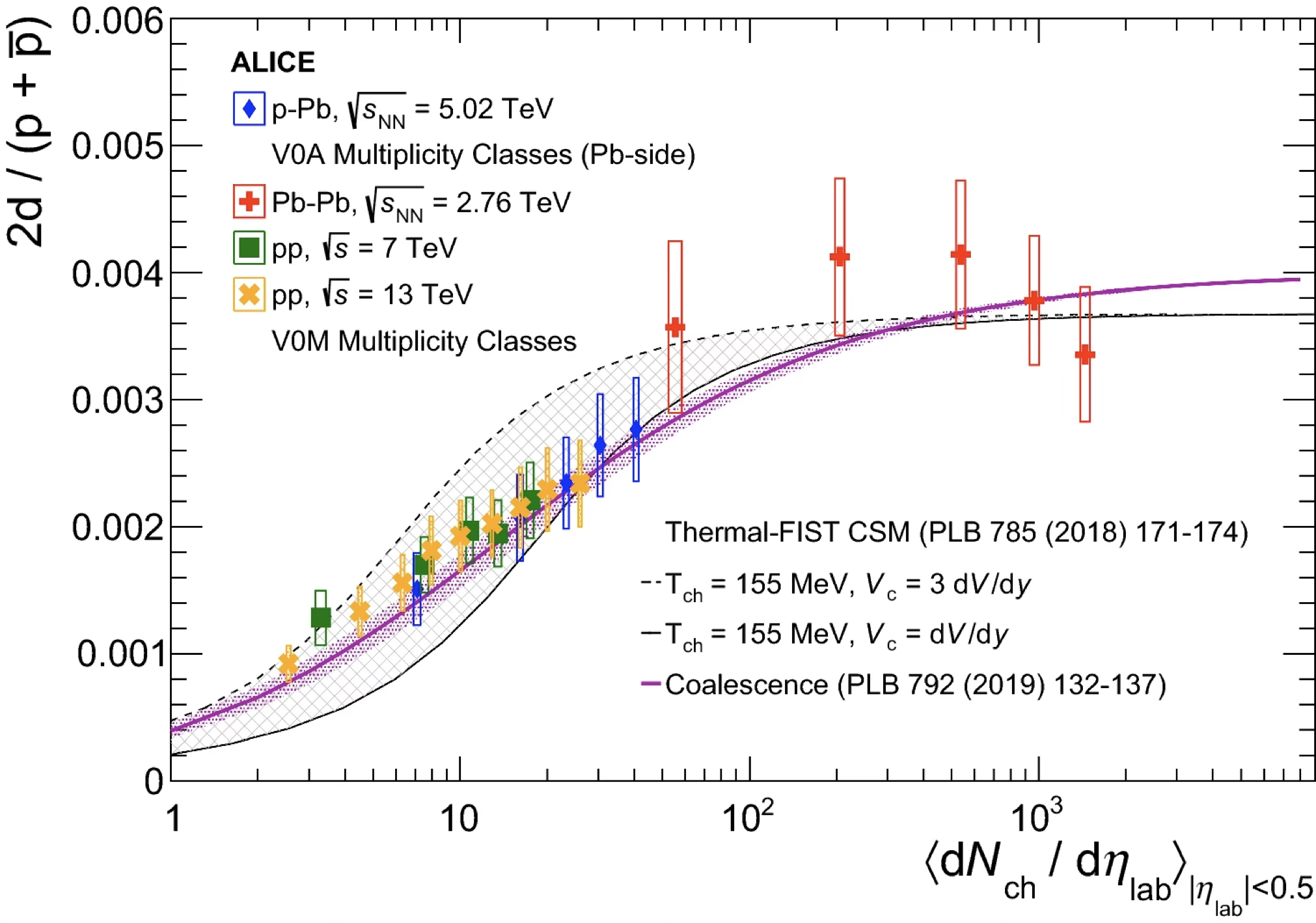}
  \caption{Deuteron-to-proton ratio as a function of charged-particle multiplicity density. Data from pp, p--Pb, and Pb--Pb collisions are compared to representative thermal and coalescence calculations \cite{alice2020}.}
  \label{fig:dtop}
\end{figure}

Two broad frameworks exist to describe the production mechanism of light nuclei. In coalescence, nuclei form late, near kinetic freeze-out, when the system has diluted enough that hadronic inelastic particle-producing interactions are no longer possible. Formation is then governed by phase-space proximity: antinucleons that are close in coordinate space and have similar momenta can merge into a bound state (see Fig.~\ref{fig:coalescence}). While intuitively and traditionally this does sound more feasible, as highlighted in Refs.~\cite{korsmeier2018,kapusta1980,sato1981,nagle1996,bellini2021,scheibl1999,mahlein2023}, experimentally it has been difficult to get an idea of when these particles are produced in colliders. In contrast, in the statistical hadronisation model (SHM), also known as the thermal model, the relative yields of particles are set earlier, at chemical freeze-out or hadronization, and follow equilibrium abundances characterized mainly by a temperature, $T$, and baryochemical potential. At the LHC, where matter--antimatter symmetry makes the relevant chemical potentials essentially vanish, a single $T = 150$--$160$ MeV describes a remarkably broad set of light-flavour ($u$, $d$, $s$ valence quarks) hadron abundances and all the light nuclei yields, as shown in Fig.~\ref{fig:shm} \cite{andronic2018,vovchenko2017,becattini2014,andronic2011}. The SHM assumes that all the particles, including the light nuclei, are formed directly at the hadronization stage following equilibrium thermal production.

This is the origin of the well-known ``Snowballs in Hell'' puzzle: how can a thermal description at $\sim155$ MeV \cite{andronic2018,vovchenko2017,becattini2014,andronic2011} succeed for objects with binding energy of a few-MeV scale? For example, the binding energy for deuterons is $\sim2.23$ MeV. The key is that the thermal model need not imply that fully formed nuclei ``survive'' inside a hot medium. Rather, it is a statement about statistical weights and the partitioning of conserved quantum numbers at hadronization, with the later evolution largely preserving the established chemistry. The real tension is therefore not that thermal models reproduce yields, but that they do so as well as dynamical coalescence pictures. This is illustrated in Fig.~\ref{fig:dtop}, which shows the deuteron-to-proton ratio as a function of event activity across pp, p--Pb and Pb--Pb collisions. Both the thermal baseline, Thermal-FIST with canonical suppression, and a coalescence calculation can describe the same smooth rise with multiplicity. The agreement could be accidental, but it may also be telling us something deeper about how QCD organizes hadron formation into composite objects, and why inclusive yields alone are an ambiguous guide to the underlying mechanism. The prediction of the relative yields by the statistical thermal models could be accidental but could also provide fundamentally new insights into QCD and the hadronization process. This is why matching integrated yields and particle ratios is not enough. The discriminating power lies in differential measurements, especially looking at transverse-momentum (\pT) dependence, multiplicity scaling, and trends from pp to p--Pb to Pb--Pb collisions. In correlation observables, this can be further probed microscopically, whether nuclei inherit nucleon-level structure or behave like species fixed at hadronization. In this sense, light-nuclei studies are not just for curiosity: they are precision tools on hadronization and the late hadronic phase, and their apparent simplicity may point to a principle of QCD matter that we have not yet fully articulated.

The ``Snowballs in Hell'' tension is often phrased as a survival paradox: how could nuclei with binding energies of only a few MeV be associated with a hadronization temperature of $T = 150$--$160$ MeV? In parallel, coalescence remains a compelling dynamical alternative in which nuclei form late from nearby nucleons in phase space, making nuclei sensitive probes of the space-time structure of the expanding system.

This is also why most widely used event generators at the LHC do not include light-nuclei production in their default settings. Popular event generators were designed primarily to model parton showers, hadronization, and the production of single hadrons across a broad range of observables; light nuclei are different because they are composite objects whose formation is expected to depend on multi-particle correlations and on the late, dilute stage of the collision evolution. Since there is still no consensus on whether nuclei emerge predominantly through late-stage coalescence or are effectively produced at hadronization in a statistical picture, including nuclei by default would require favouring a specific and still debated microscopic mechanism, along with additional modelling choices and parameters. In practice, when nuclei are included, they often appear as optional add-ons or afterburners; for example, deuteron formation via a coalescence module can be enabled in PYTHIA rather than being on by default. This reflects that the field treats light nuclei not as a settled part of the baseline event description, but as a sensitive diagnostic of hadronization and freeze-out dynamics.

\begin{figure}[t]
  \centering
  \includegraphics[width=0.98\columnwidth]{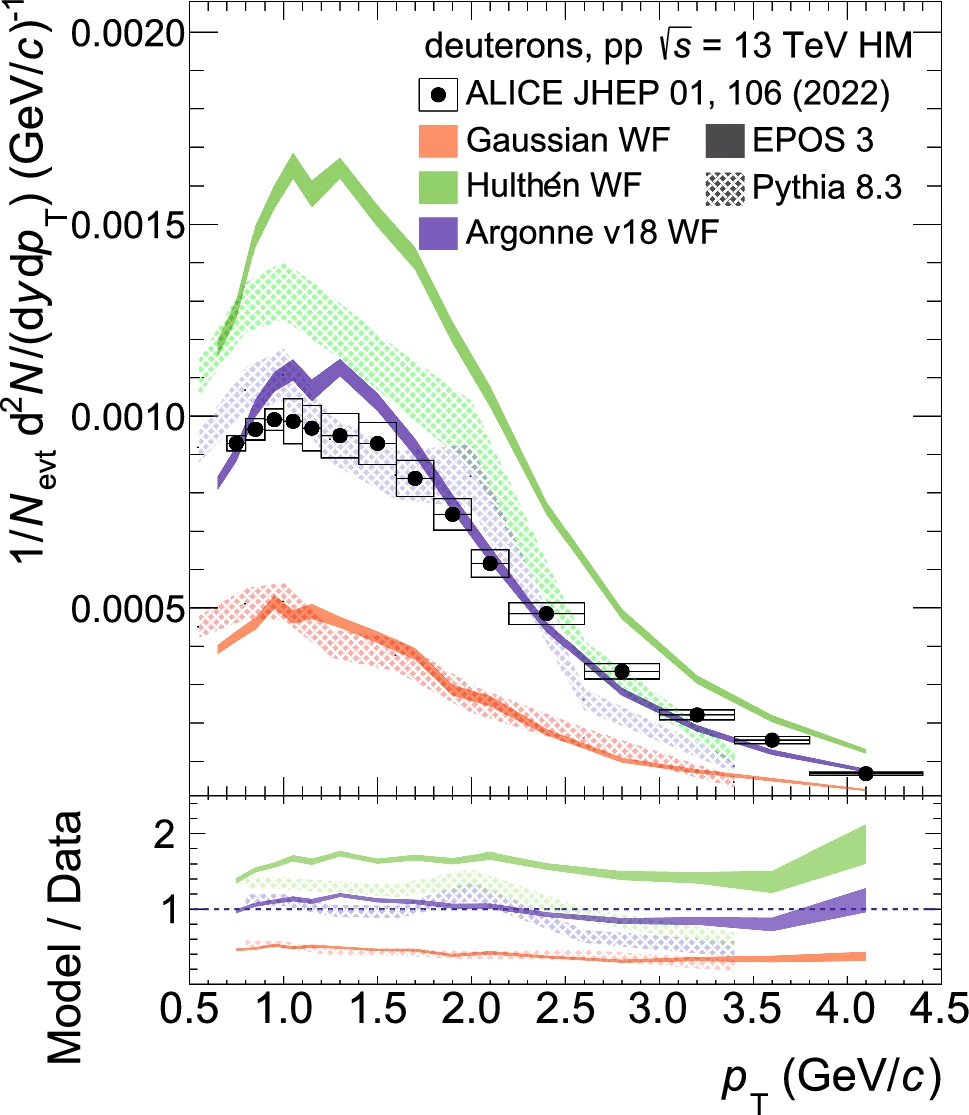}
  \caption{Deuteron \pT{} spectrum in high-multiplicity pp collisions at \sqrts{} = 13 TeV measured by ALICE, compared to event-generator-based coalescence calculations using a Wigner-function afterburner \cite{mahlein2023}.}
  \label{fig:pt-spectrum}
\end{figure}

A recent coalescence study \cite{mahlein2023} shows that once one goes beyond schematic momentum-space coalescence, the predicted deuteron spectra become sensitive to genuinely late-stage, multi-particle physics: one needs an event-by-event afterburner based on the Wigner-function formalism,\footnote{The Wigner function provides a phase-space representation of the deuteron wavefunction, allowing the coalescence probability to depend simultaneously on the relative momentum and relative distance of the proton-neutron pair.} realistic space-momentum correlations among nucleons, and crucially an emitting-source size anchored to femtoscopic measurements in the same collision system. The work demonstrates that ``native'' sources from standard generators do not reproduce the measured transverse-mass dependence of the baryon source radius and must be calibrated to data. With these constraints in place, the deuteron spectra show a clear dependence on the assumed deuteron wavefunction. This sensitivity is illustrated in Fig.~\ref{fig:pt-spectrum}, where the measured \pT-differential deuteron yield is compared to coalescence predictions using different wavefunctions, and a realistic choice, Argonne v18,\footnote{Argonne v18 is a realistic deuteron wavefunction based on nucleon-nucleon scattering constraints \cite{mahlein2023}.} gives the best agreement with the ALICE data.

\section*{Observation of nuclei from resonance decay}

\begin{figure}[t]
  \centering
  \includegraphics[width=0.9\columnwidth]{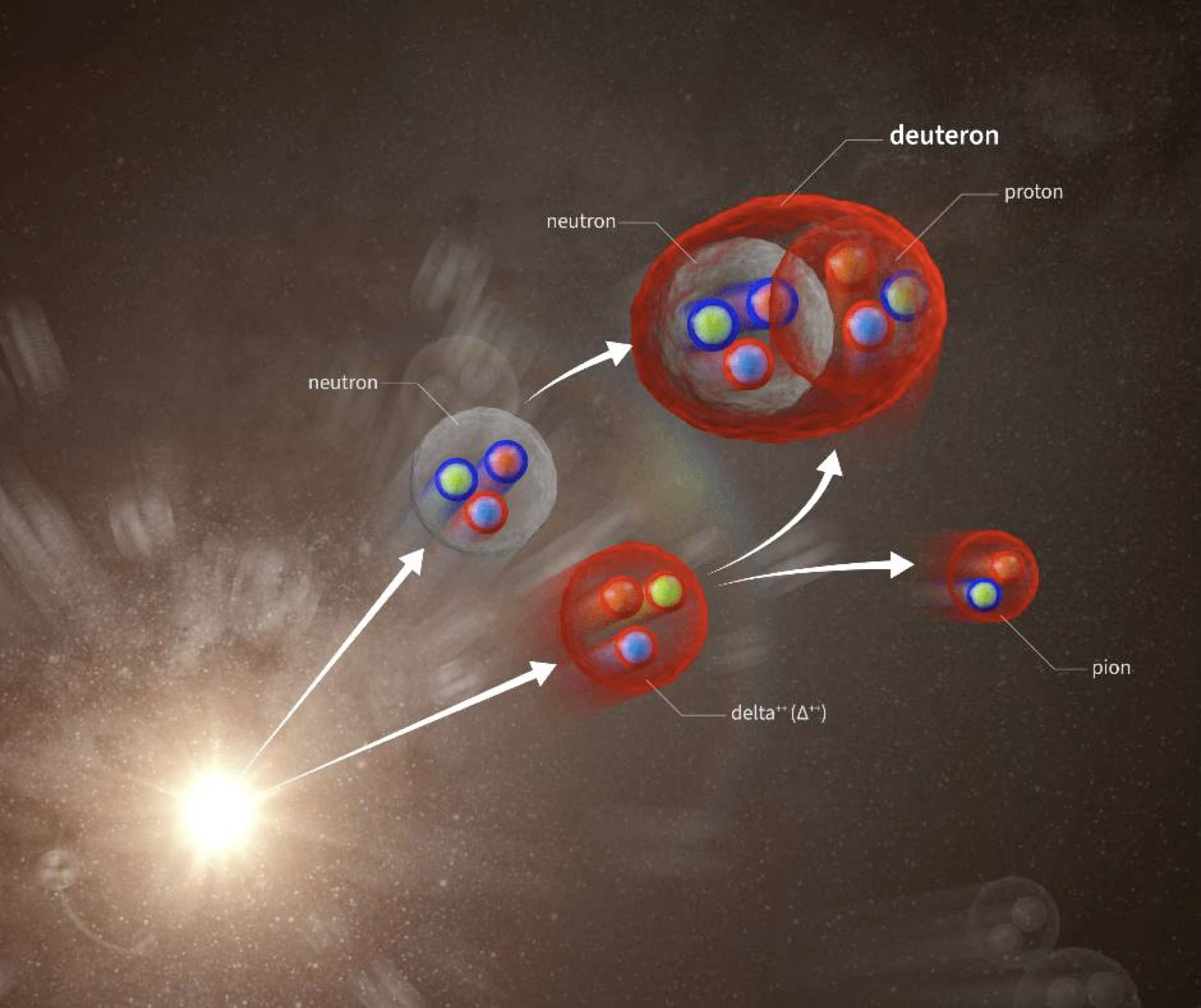}
  \caption{Schematic illustration of resonance-fed deuteron formation in high-energy collisions \cite{fabbietti2025}. Short-lived baryon resonances decay into nucleons and pions; as the system expands and dilutes, nearby proton-neutron pairs can coalesce into a deuteron, while the accompanying pion can retain a residual momentum correlation that provides an experimental handle on this microscopic pathway.}
  \label{fig:resonance-fed}
\end{figure}

An important recent step came from the ALICE Collaboration in a Nature article \cite{alice2025nature} reporting the first direct evidence for a specific microscopic pathway to deuteron and antideuteron formation in pp collisions. By analysing deuteron-pion momentum correlations, ALICE isolates a distinctive residual-correlation pattern that is difficult to reproduce with a purely thermal scenario and instead points to a fusion-like formation process occurring after resonance decays. The key observation is that the measured deuteron-pion correlation function carries the imprint of pion-nucleon pairs originating from the same $\Delta(1232)$ decay, implying that the nucleon that later ends up inside the deuteron was produced in a strong resonance decay and remained correlated with the accompanying pion. Interpreting the data with this decomposition, ALICE reports largely model-independent evidence that a dominant fraction, about 90\%, of the observed (anti)deuterons are produced in nuclear reactions following the decay of short-lived resonances such as the $\Delta(1232)$. This reframes the ``Snowballs in Hell'' puzzle: instead of requiring fragile nuclei to survive an early hot stage, the data in pp collisions support a picture where resonance decays supply correlated nucleons and deuterons form later as the system dilutes, as illustrated in Fig.~\ref{fig:resonance-fed}. Beyond closing a long-standing understanding in collider nucleosynthesis, the result also feeds into modelling antinuclei production in cosmic rays and in potential dark-matter decay scenarios.

\section*{Way forward}

\begin{figure}[t]
  \centering
  \includegraphics[width=0.98\columnwidth]{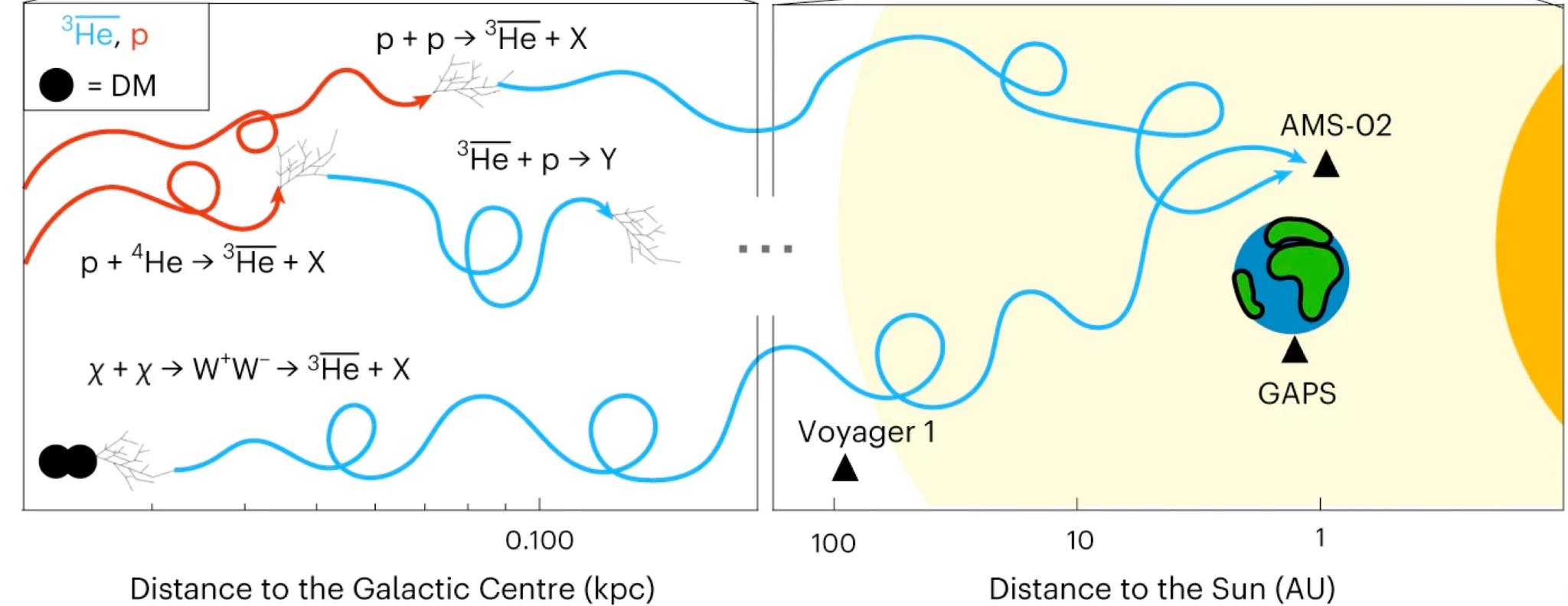}
  \caption{Schematic view of antihelium propagation from its production sites in the Galaxy to near-Earth detectors. Along the journey from the Galactic Centre region to the Solar System, light nuclei can be produced in cosmic-ray interactions (secondary background) or in exotic scenarios such as dark-matter annihilation and decay and are then shaped by transport through the interstellar medium and solar modulation in the heliosphere \cite{alice2023np}.}
  \label{fig:antihelium}
\end{figure}

The next major step is to move from asking whether light nuclei are produced, to asking how they are produced event-by-event and particle-by-particle. This is where the upcoming LHC data will be important. In recently concluded Run 3 at the LHC, ALICE has been operating with a substantially upgraded detector and continuous readout, designed to handle much larger heavy-ion collision rates than before, reaching Pb--Pb interaction rates of about 50 kHz. This means that rare objects such as deuterons, antideuterons and heavier antinuclei can be studied not only through their total yields, but through detailed correlations with the other particles produced in the same collision. Run 4 will push this further as ALICE has approved upgrades such as ITS3 and FoCal for installation during the next long shutdown, improving precision tracking close to the collision point and extending the experimental reach in new directions.

A particularly promising route is through balance functions. Instead of only counting how many deuterons are produced, balance functions ask where the corresponding balancing quantum numbers go. If a deuteron carries baryon number, charge and isospin, then the particles balancing these quantum numbers must also appear somewhere in the same event. By measuring deuteron-triggered and proton-triggered balance functions, one can test whether nuclei are assembled late from nearby nucleons, as expected in coalescence, or whether their abundances and correlations are already largely fixed near hadronization, as in the thermal picture \cite{tripathy2026epjc}. The crucial point is that the two pictures may give similar integrated yields, but they need not give the same correlation patterns. In particular, the dependence on transverse momentum, event multiplicity and angular separation can become a microscope for the space-time origin of light nuclei.

One may even try to look for the parentage of nuclei more directly. A possible idea is to use short-lived resonances such as \Lambdastar{} as a controlled probe of the hadronic stage \cite{tripathy2026arxiv}. Since such resonances decay into correlated hadrons and carry well-defined quantum numbers, their correlation with protons, deuterons or other light nuclei may reveal whether the final nuclei remember the decay history of the system. If Run 3 and Run 4 data can show that light nuclei are preferentially connected to nearby baryons, resonance decay products or specific balancing partners, then the ``Snowballs in Hell'' puzzle will no longer be only a question of survival. It will become a measurable story of how simple nuclear objects emerge from the most extreme form of QCD matter created in the laboratory.

\section*{Why light nuclei matter for indirect dark-matter searches}

What makes this debate matter beyond heavy-ion physics is that the same light nuclei are also important in indirect dark-matter searches with cosmic rays. Low-energy antinuclei, especially antideuterons, and potentially antihelium, are crucial as conventional astrophysical production is expected to be strongly suppressed in the low-energy region where dark-matter scenarios predict comparatively enhanced signals \cite{hailey2009,doetinchem2020,poulin2019,ting2016}. The challenge is that turning a possible detection into a constraint on dark matter depends not only on production, but also on propagation. Antinuclei can be absorbed or fragmented while travelling through the interstellar medium and heliosphere. A schematic view of antihelium propagation is shown in Fig.~\ref{fig:antihelium}.

This is where the ALICE measurement becomes valuable. Using \HeThree{} produced at the LHC, ALICE extracted the inelastic interaction cross section of \HeThree{} with matter by treating the detector material itself as a target and then used that result to quantify how ``transparent'' the Milky Way is to such antinuclei during propagation. In a benchmark calculation, ALICE found a Galactic transparency of order 50\% for \HeThree{} originating from a dark-matter source, while for conventional cosmic-ray production it ranges roughly from 25\% to 90\%, depending on energy \cite{alice2023np}. The key point is not that the LHC tests dark matter directly, but that collider measurements can remove a major nuclear-physics uncertainty in the predicted flux at Earth, tightening how experiments such as AMS-02 and the GAPS experiment \cite{hailey2009,doetinchem2020,poulin2019,ting2016} interpret any future antinucleus candidates.

\section*{Summary and outlook}

Light nuclei at the LHC remain interesting as they expose a deep gap in our microscopic understanding of hadronization. The core of the ``Snowballs in Hell'' puzzle is not that thermal models can reproduce light-nucleus yields with a hadronization temperature of $T \sim 155$ MeV despite MeV binding energies, but that they do so as well as dynamical coalescence pictures. The last decade of measurements has made it clear that inclusive ratios and integrated yields are therefore not decisive: both frameworks have been successful in describing the same global trends across pp, p--Pb and Pb--Pb collisions. Progress has instead come from observables that are sensitive to space-time structure and multi-particle correlations. On the theory side, realistic coalescence implementations now show that deuteron spectra depend on the emitting source size, constrained by femtoscopy, resonance feed-down, and even on the deuteron wavefunction, exactly the kind of late-stage physics that standard event generators were not originally designed to include by default. On the experimental side, the recent ALICE Nature result on pion-deuteron correlations provides the first correlation-based evidence for resonance-fed deuteron formation, strongly suggesting that a large fraction of (anti)deuterons are formed in nuclear reactions following short-lived baryon-resonance decays.

The outlook is therefore clear: the next breakthroughs will come from moving beyond integrated yields and by using differential and correlation measurements as a microscope. Run 3 at the LHC already provides unprecedented statistics and precision for identified-particle studies, enabling systematic scans in transverse momentum, multiplicity, and system size where the thermal and coalescence pictures need not make the same predictions. Balance functions and other quantum-number correlation observables are well suited to distinguish scenarios where nuclei inherit nearby nucleon structure, as expected for coalescence, from scenarios where correlations are largely washed out and yields are set statistically near hadronization. Resonance-tagging strategies, especially those exploiting clean hadronic decays and controlled lifetimes, offer an additional handle on parentage by asking whether nuclei retain measurable memory of decay nucleons from specific sources. As these measurements mature, they will also sharpen the astrophysics connection of cosmic-ray searches for antinuclei, which rely on credible predictions for secondary backgrounds, and collider data are uniquely positioned to constrain the hadronic production and propagation ingredients that presently limit the interpretation of any future antinuclei candidates in terms of dark-matter scenarios. In short, light nuclei are no longer just a curiosity; they are becoming precision tools for pinning down how QCD matter turns into composite objects and for advancing cosmic-ray studies and indirect dark-matter searches.

\section*{Acknowledgements}

S.T. acknowledges the funding received from the European Union's Horizon Europe research and innovation programme under the Marie Sklodowska-Curie grant agreement No. 101149298. R.S. gratefully acknowledges the DAE-DST, Government of India funding under the mega-science project ``Indian participation in the ALICE experiment at CERN'', bearing Project No. SR/MF/PS-02/2021-IITI (E-37123).

\vspace{1em}
\noindent\rule{\columnwidth}{0.4pt}

\noindent\textit{Dr. Sushanta Tripathy is a Marie-Curie Fellow at the University of Lund, Sweden, and Professor Raghunath Sahoo is an Institute Chair Professor at the Indian Institute of Technology Indore, India.}

\vspace{0.4em}
\noindent\textit{Correspondence: \href{mailto:Raghunath.Sahoo@cern.ch}{Raghunath.Sahoo@cern.ch}}

\end{document}